 \documentclass{SCIS2024}

\usepackage{graphicx}
\usepackage{picins}
\usepackage{amsmath}
\usepackage{lmodern}
\usepackage{amssymb}
\usepackage{subfig}
\usepackage{lineno}
\usepackage{makecell}
\usepackage{indentfirst}
\usepackage{bm}
\usepackage{color}
\usepackage{cite}
\usepackage{epstopdf}
\usepackage{amssymb}
\usepackage{xcolor}
\usepackage{tikz}
\usepackage{mathtools}
\usepackage{amsthm}
\usepackage{mathrsfs}
\usepackage{verbatim}
\usepackage{supertabular} 

\usepackage{float}
\usepackage{hyperref}
\hypersetup{
	colorlinks=true,
	linkcolor=blue,
	filecolor=magenta,      
	urlcolor=cyan,
	pdfborderstyle={/S/U/W 1} 
}
\usepackage[algo2e]{algorithm2e}
\begin{document}
	\ArticleType{RESEARCH PAPER}
	\Year{2024}
	\Month{}
	\Vol{}
	\No{}
	\DOI{}
	\ArtNo{}
	\ReceiveDate{}
	\ReviseDate{}
	\AcceptDate{}
	\OnlineDate{}
	\title{Near-Field Channel Estimation for Extremely Large-Scale Terahertz Communications}{Near-Field Channel Estimation for Extremely Large-Scale Terahertz Communications}	
	\author[1]{Songjie Yang}{}
	\author[1]{Yizhou Peng}{}
	\author[1]{Wanting Lyu}{}
	\author[2]{Ya Li}{}
	\author[2]{Hongjun He}{}
	\author[1]{\\Zhongpei Zhang}{{zhangzp@uestc.edu.cn}}
	\author[3]{Chau Yuen}{}
	
	\AuthorMark{Songjie Yang}
	
	\AuthorCitation{Songjie Yang, Yizhou Peng, Wanting Lyu, et al}

	
	\address[1]{ the National
		Key Laboratory of Wireless Communications, \\ University of Electronic Science and Technology of China, Chengdu {\rm 611731}, China}
	\address[2]{ the Future Research
		Laboratory, China Mobile Research Institute, Beijing, China}
	\address[3]{ the School of Electrical and Electronics Engineering,
		Nanyang Technological University}
	
	\abstract{Future Terahertz communications exhibit significant potential in accommodating ultra-high-rate services. Employing extremely large-scale array antennas is a key approach to realize this potential, as they can harness substantial beamforming gains to overcome the severe path loss and leverage the electromagnetic advantages in the near field.
	This paper proposes novel estimation methods designed to enhance efficiency in Terahertz widely-spaced multi-subarray (WSMS) systems.
	 Initially, we introduce three sparse channel representation methods: polar-domain representation (PD-R), multi-angular-domain representation (MAD-R), and two-dimensional polar-angular-domain representation (2D-PAD-R). Each method is meticulously developed for near-field WSMS channels, capitalizing on their sparsity characteristics. Building on this, we propose four estimation frameworks using the sparse recovery theory: polar-domain estimation (PD-E), multi-angular-domain estimation (MAD-E), two-stage polar-angular-domain estimation (TS-PAD-E), and two-dimensional polar-angular-domain estimation (2D-PAD-E). Particularly, 2D-PAD-E, integrating a 2D dictionary process, and TS-PAD-E, with its sequential approach to angle and distance estimation, stand out as particularly effective for near-field angle-distance estimation, enabling decoupled calculation of these parameters.
	  Overall, these frameworks provide versatile and efficient solutions for WSMS channel estimation, balancing low complexity with high-performance outcomes. Additionally, they represent a fresh perspective on near-field signal processing.}
	

	\maketitle

\section{Introduction}  
To cater to a diverse range of future applications, there is a growing demand for high-rate communication and high-resolution sensing. In pursuit of this objective,
various emergying technologies are provital, such as high-frequency communications that provide large widebandth \cite{HF,THZ}, reconfigurable intelligent surfaces (RISs) that improve the wireless channel condition \cite{RIS1,RIS2}, massive multi-input multi-output (MIMO) that provide strong spatial multiplexing \cite{MMIMO}, and holographic MIMO \cite{J1,J2}. The future potential of these promising technologies lies in their ability to intricately intertwine and mutually enhance one another. In addition, they can undergo advanced evolution and enhancements through the utilization of extremly large-scale (XL)-array technologies, exemplified by recent research on XL-RIS \cite{XL-RIS1,NCE2} and XL-MIMO \cite{XL-MIMO}. Notably, XL-array technology within the THz frequency range is of paramount importance.

XL-arrays produce highly directional beam patterns, thereby enhancing spatial resolution and beamforming gain significantly. However, it's important to note that this approach also alters the electromagnetic propagation characteristics compared to traditional communication methods. 
\textcolor{red}{As the array aperture increases, the enlarged Rayleigh distance, which marks the boundary between the near- and far-field regions \cite{E0,E1}}, must be given significant attention within the typical communication coverage range due to its substantial impact.  Consequently, it is imperative to delve into near-field signal processing. For instance, near-field localization and channel estimation warrant attention due to the altered spatial characteristics. This has garnered the interest of some researchers and has been examined in diverse contexts \cite{NCE1,NCE2,NCE3,NCE4,UPACE,UPACE2,SE1,SE2,SE3,SE4,SC,nf-loc,loc-syn,ss1}.
In \cite{NCE1}, the polar-domain (PD) codebook/dictionary was proposed to characterize the near-field beamspace for channel estimation, which demonstrated that uniform distance reciprocal sampling outperformed uniform distance sampling when employing greedy recovery algorithms. \cite{NCE3} and \cite{NCE4} investigated wideband channel estimation, exploring the impact of near-field beam squint effect on angle-distance parameter estimation and suggesting joint angle-distance estimation methods across all subcarriers. Moreover, \cite{UPACE} and \cite{UPACE2} investigated near-field parameter estimation with uniform planar arrays, introducing a decoupled estimation strategy to simplify the angle and distance estimation process, significantly reducing computational complexity and providing promising avenues for near-field multi-parameter estimation.
 
All of these methods center around processing the entire array for near-field signal processing. Additionally, certain subarray-based approaches leverage the characteristic that each subarray adheres to the planar wavefront assumption. Examples include \cite{loc-syn} for joint localization and synchronization, and \cite{ss1} for channel estimation. Expanding on this notion, particularly notable is the research focused on widely-spaced multi-subarray (WSMS) scenarios \cite{SE1,SE2,SE3,SE4}.
The WSMS layout becomes promising for Teraherz (THz) array antennas \cite{SE1,SE2,SE3,SE4,WS3,WS4,WS5,WS-MIMO1,WS-MIMO2,CRB,CFN}, due to 1) array scalability and flexity, 2) manufacturing feasiblity, 3)
simplified circuitry and signal processing, and 4) size and weight considerations.  By leveraging the spatial structure of WSMSs and the sparsity of THz channels \cite{T1,T2}, researchers adopted the hybrid spherical and planar wavefront (HSPW) assumption \cite{WS-MIMO2,WS-MIMO1}. Here, the planar wavefront applies to intra-subarray interactions, while the spherical wavefront pertains to inter-subarray interactions, simplifying signal processing procedures.  This approach can also be referred to as cross-field model, where far-field and near-field models are employed for antenna-level and subarray-level responses, respectively. In this context, multi-subarray beamforming and capacity analysis have been investigated. Particularly,
with the same number of antennas, \cite{WS-MIMO1} proved that WSMSs could provid stronger multiplexing cabilities 
than the uniform XL-arrays by increasing the inter-subarray spacing to enlarge the near-field effect. 
  In \cite{SE1}, spatial parameters were estimated using a parallel approach. This approach treats each subarray independently, aiming to reduce the complexity associated with estimating the channel of the entire array. In \cite{SE2}, researchers adopted a sequential approach for subarray estimation. They estimated parameters for one subarray while keeping others fixed, iteratively repeating this process.
Additionally, in \cite{SE3} and \cite{SE4}, an alternative subarray estimation method was introduced. It initially estimated spatial parameters for the first subarray and subsequently utilized these estimates to create a reduced dictionary for the remaining subarrays. This method effectively reduces dictionary redundancy, resulting in enhanced parameter recovery. In \cite{CRB}, we derived closed-form Cram\'er-Rao bounds for near-field localization with WSMSs, highlighting their potential for angle and range estimation. Additionally, we illustrated the effectiveness of employing the cross-field model for WSMSs.

Essentially, the works \cite{SE1, SE2, SE3, SE4} all fall within a multi-angular-domain estimation framework, aligning each subarray to an angle beam to approximate an angle-distance beam, albeit utilizing distinct methodologies. While these approaches offer practicality for low-complexity implementation in WSMS channel estimation, they share a common limitation: the oversight of the distance-weighted angular array response in the cross-field model. This oversight hampers their ability to derive the distance parameter, limiting their applicability in certain scenarios. This limitation has spurred us to propose innovative estimation frameworks.

One approach is to extend the PD dictionary \cite{NCE1} to the WSMS case for angle-distance estimation. However, this approach may encounter high complexity and does not fully exploit the properties of WSMSs. In this paper\footnote{The source code is available at \url{https://github.com/YyangSJ/WSMS-NF-CE} for readers' study.}, we aim to propose more efficient estimation frameworks for THz WSMSs by leveraging the cross-field model. The main contributions are summarized as follows.
\begin{itemize}
	\item \textcolor{red}{Drawing upon the inherent properties of WSMS structures, we can employ approximations of the exact near-field model to design efficient methods for near-field channel estimation.}
	We begin by proposing three sparse channel representation approaches for WSMS channels, namely  polar-domain representation (PD-R), multi-angular-domain representation (MAD-R), 2D polar-angular-domain representation (2D-PAD-R), respectively. Specifically, PD-R harnesses polar-domain sparsity within the WSMS channel by employing a sparse vector and a PD dictionary to characterize the channel. MAD-R views each subarray as a planar wavefront, adopting the conventional far-field angular-domain channel representation for each subarray. In contrast, 2D-PAD-R embraces the HSPW assumption to construct a near-field channel representation, underpinned by a sparse matrix.
	\item We discuss a direct framework to solve the WSMS channel estimation problem based on PD-R, referred to as polar-domain estimation (PD-E). This framework capitalizes on the established compressive sensing technique to extract the critical angle and distance parameters required for channel reconstruction. Nevertheless, in recognition of its heightened complexity, we introduce an alternative solution: multi-angular-domain estimation (MAD-E), which conducts angular-domain recovery for each small-scale subarray, pursuing a low-complexity approach.
	 Moreover, building upon the foundation of 2D-PAD-R, we propose a two-stage polar-angular-domain estimation (TS-PAD-E) framework that unifies all subarrays to estimate angles through a multiple measurement vector (MMV) problem. Subsequently, it extracts distances from the estimated MMVs. To address the limitation of this framework, namely, the accumulation of estimation errors inherent in the two-stage process, we further propose a 2D polar-angular-domain estimation (2D-PAD-E) framework that performs a 2D-atom recovery procedure.
	 \item Building on the proposed estimation frameworks, we incorporate the commonly used compressive sensing algorithm, Orthogonal Matching Pursuit (OMP), into PD-OMP, MAD-OMP, TS-PAD-OMP, and 2D-PAD-OMP, facilitating the greedy recovery of angle and distance parameters.
	\item Finally, measurement matrix optimization in WSMSs is considered for estimation enhancement.
	
\end{itemize}

The rest of this paper is organized as follows: Section \ref{system} describes the signal model and the channel model for WSMS systems.
Section \ref{DR} discusses three dictionary representation approaches for near-field channels, including PD-R, MAD-R, and 2D-PAD-R. Section \ref{EF} proposes four estimation frameworks leveraging the three channel representation approaches. Based on the proposed estimation frameworks, Section \ref{EOMP} employs the OMP algorithm within the four estimation frameworks to perform the greedy recovery of angle and distance parameters.
Section \ref{MMO} optimizes the measurement matrix to enhance the estimation performance.
 Section \ref{SR} conducts several experiments to demonstrate our proposed methods' effectiveness. Finally, Section \ref{Con} concludes this paper and provides some potential directions.

{\emph {Notations}}:
${\left(  \cdot  \right)}^{ *}$, ${\left(  \cdot  \right)}^{ T}$, ${\left(  \cdot  \right)}^{ H}$, and $\left(\cdot\right)^{-1}$ denote conjugate, transpose, conjugate transpose, and inverse, respectively. $  \mathbf{A}^{\dagger}$ represents the pseudo inverse such that $  \mathbf{A} ^{\dagger}=(\mathbf{A}^H\mathbf{A})^{-1}\mathbf{A}^H$.  $\Vert\cdot\Vert_0$,
 $\Vert\cdot\Vert_2$, and  $\Vert\cdot\Vert_F$ represent $\ell_0$ norm, $\ell_2$ norm, and Frobenius norm, respectively.  $\vert\cdot\vert$ denotes the modulus. Furthermore, $\otimes$ is the Kronecker product.  $[\mathbf{A}]_{i,:}$ and $[\mathbf{A}]_{:,j}$ denote the $i$-the row and $j$-th column of $\mathbf{A}$, respectively. ${\rm blkdiag}{\cdot}$ denotes the block-diagonal operator.
  Finally, $\mathcal{CN}(\mathbf{a},\mathbf{A})$ is the complex Gaussian distribution with mean $\mathbf{a}$ and covariance matrix $\mathbf{A}$.

\begin{figure*}
	\centering
	\includegraphics[width = 0.75\textwidth]{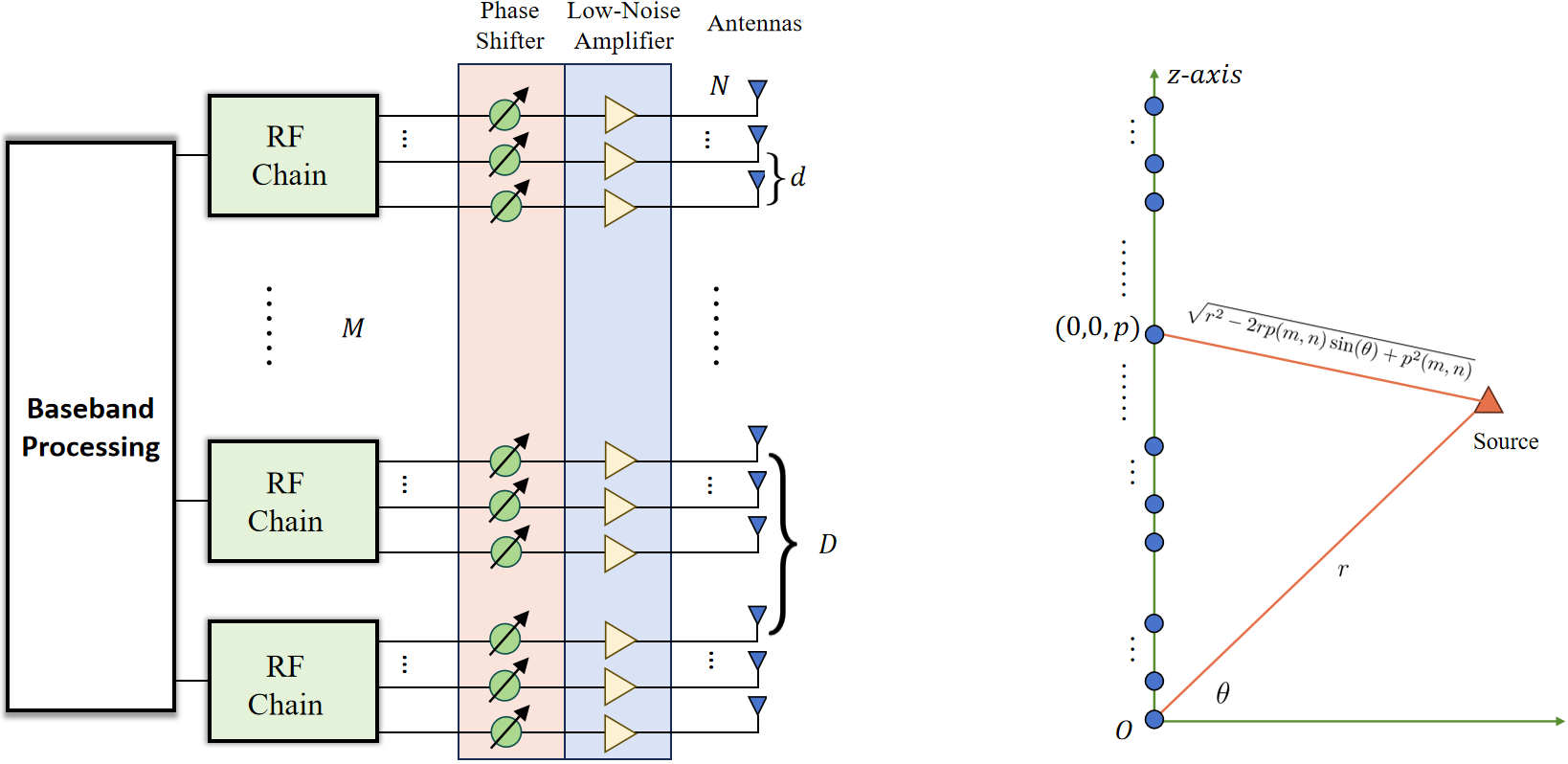}
	\caption{WSMS Structure (Left) and Coordinate System (Right)}
	\label{sys_fig}
\end{figure*}
\section{System Model}\label{system}
This paper focuses on a near-field uplink narrowband training scenario, wherein the base station utilizes a WSMS structure. This system comprises $M$ radio frequency (RF) chains, with each chain linked to $N$ antenna elements, as depicted in Fig. \ref{sys_fig}. The user is equipped with a single antenna. For clarity, we consider a uniform linear layout\footnote{Complex cases like MIMO and uniform planar arrays (UPAs) are discussed in the final section.}, wherein the inter-antenna spacing is denoted as $d$ and the inter-subarray spacing as $D$.
With the use of an orthogonal pilot sequence, multi-user channel estimation can be treated as several parallelizable problems. Without losing generality, we discuss the pilot signal model for one arbitrary user.
 
\subsection{Uplink Training}
As one user successfully sends the pilot signal,
the received signal at the $m$-the RF chain for the $q$-th pilot symbol ($m\in\{1,\cdots,M\}$, $q\in\{1,\cdots,Q\}$) is given by
\begin{equation}
	{y}_{m,q}=\mathbf{w}_{m,q}^H\mathbf{h}_m{s}_q+\mathbf{w}_{m,q}^H\mathbf{n}_{m,q},
\end{equation}
where ${s}_q$, set to $1$, is the $q$-th pilot signal, $\mathbf{h}_m\in\mathbb{C}^{N\times 1}$ denotes the wireless channel between the $m$-the RF chain and the user, $\mathbf{w}_{m,q}\in\mathbb{C}^{N\times 1}$ is the combing vector for sampling the channel,
and $\mathbf{n}_{m,q}$ is the independent and identically distributed additive white Gaussian noise vector following the distribution $\mathcal{C} \mathcal{N}\left(0, \sigma_n^{2}\mathbf{I}\right)$.

By collecting the $Q$ pilots,   the $m$-th RF chain's received signal $\mathbf{y}_m=[y_{m,1},\cdots,y_{m,Q}]^T\in\mathbb{C}^{Q\times 1}$ is expressed as
\begin{equation}\label{y}
\mathbf{y}_m=\mathbf{W}_m^H\mathbf{h}_m+\overline{\mathbf{n}}_m
\end{equation}
where $\mathbf{W}_m=[\mathbf{w}_{m,1},\cdots,\mathbf{w}_{m,Q}]\in\mathbb{C}^{N\times Q}$, and $\overline{\mathbf{n}}_m=[\mathbf{w}_{m,1}^H\mathbf{n}_{m,1},\cdots,\mathbf{w}_{m,Q}^H\mathbf{n}_{m,Q}]^T\in\mathbb{C}^{Q\times 1}$.  In this work, we assume all subarrays share the same measurement matrix, i.e., $\mathbf{W}_1=\mathbf{W}_2=\cdots=\mathbf{W}_M=\mathbf{W}$.

By collecting all signals from the $M$ RF chains, yielding
	  \begin{equation}\label{y_tilde}
	\widetilde{\mathbf{y}}=\widetilde{\mathbf{W}}^H\widetilde{\mathbf{h}}+\widetilde{\mathbf{n}}, 
\end{equation}
where $\widetilde{\mathbf{y}}\triangleq[\mathbf{y}_1^T,\cdots,\mathbf{y}_M^T]^T\in\mathbb{C}^{MN\times 1}$, $\widetilde{\mathbf{W}}\triangleq{\rm blkdiag}\{\mathbf{W},\mathbf{W},\cdots,\mathbf{W}\}\in\mathbb{C}^{MN\times MQ}$, and $\widetilde{\mathbf{n}}\triangleq [\overline{\mathbf{n}}_1^T,\cdots,\overline{\mathbf{n}}_M^T]\in\mathbb{C}^{MN\times 1}$.

\subsection{Channel Model}
Next, we specify the physical channel which can characterize
the geometrical structure and limited scattering nature. The channel $\widetilde{\mathbf{h}}=[\mathbf{h}_1^T,\cdots,\mathbf{h}_M^T]^T\in\mathbb{C}^{MN\times1}$ is written as
\begin{equation}\label{H}
	\widetilde{\mathbf{h}}=\sqrt{\frac{MN}{L}}\sum_{l=1}^{L}z_l\mathbf{g}(\theta_l,r_l),
\end{equation}
where $L$ is the number of channel paths, $z_l$ denotes the complex gain of the $l$-th path, and $\{\theta_l\}_{l=1}^L$ and $\{r_l\}_{l=1}^L$ denote the angle and distance parameters. $\mathbf{g}(\theta_{l},r_l)\in\mathbb{C}^{MN\times 1}$ represents the array response, following 
\begin{equation}\label{g}
	\begin{aligned} 	\mathbf{g}(\theta,r)\triangleq	  \frac{1}{\sqrt{MN}}
		\left[ e^{-j\frac{2\pi}{\lambda} (r^{(p(1,1))}-r)},\cdots,e^{-j\frac{2\pi}{\lambda}(r^{(p(m,n))}-r)},\cdots,e^{-j\frac{2\pi}{\lambda} (r^{(p(M,N))}-r)} \right]^T,
	\end{aligned}
\end{equation}
where $p(m,n)\triangleq (m-1)D+(n-1)d,m\in\{1,\cdots,M\},n\in\{1,\cdots,N\}$. With the array deployed along the $z$-axis, shown in Fig. \ref{sys_fig}, $r^{(p(m,n))}$ is given by 
\begin{equation}\label{rd}
	\begin{aligned}
		r^{(p(m,n))}  &\triangleq\left((r\cos(\theta))^2
		+(r\sin(\theta)-p(m,n))^2 \right)^{\frac{1}{2}} \\
		&= \sqrt{r^2-2rp(m,n)\sin(\theta)+p^2(m,n)} \\
		& \overset{(a)}{\approx} r- p(m,n)\cos(\theta) +\frac{p^2(m,n)}{2r}(1-\cos^2(\theta)),
	\end{aligned}
\end{equation}
where $(a)$ holds due to $\sqrt{1+x}\approx1+\frac{x}{2}-\frac{x^2}{8}$.
 
\begin{figure*}
	\centering
	\includegraphics[width = 0.62\textwidth]{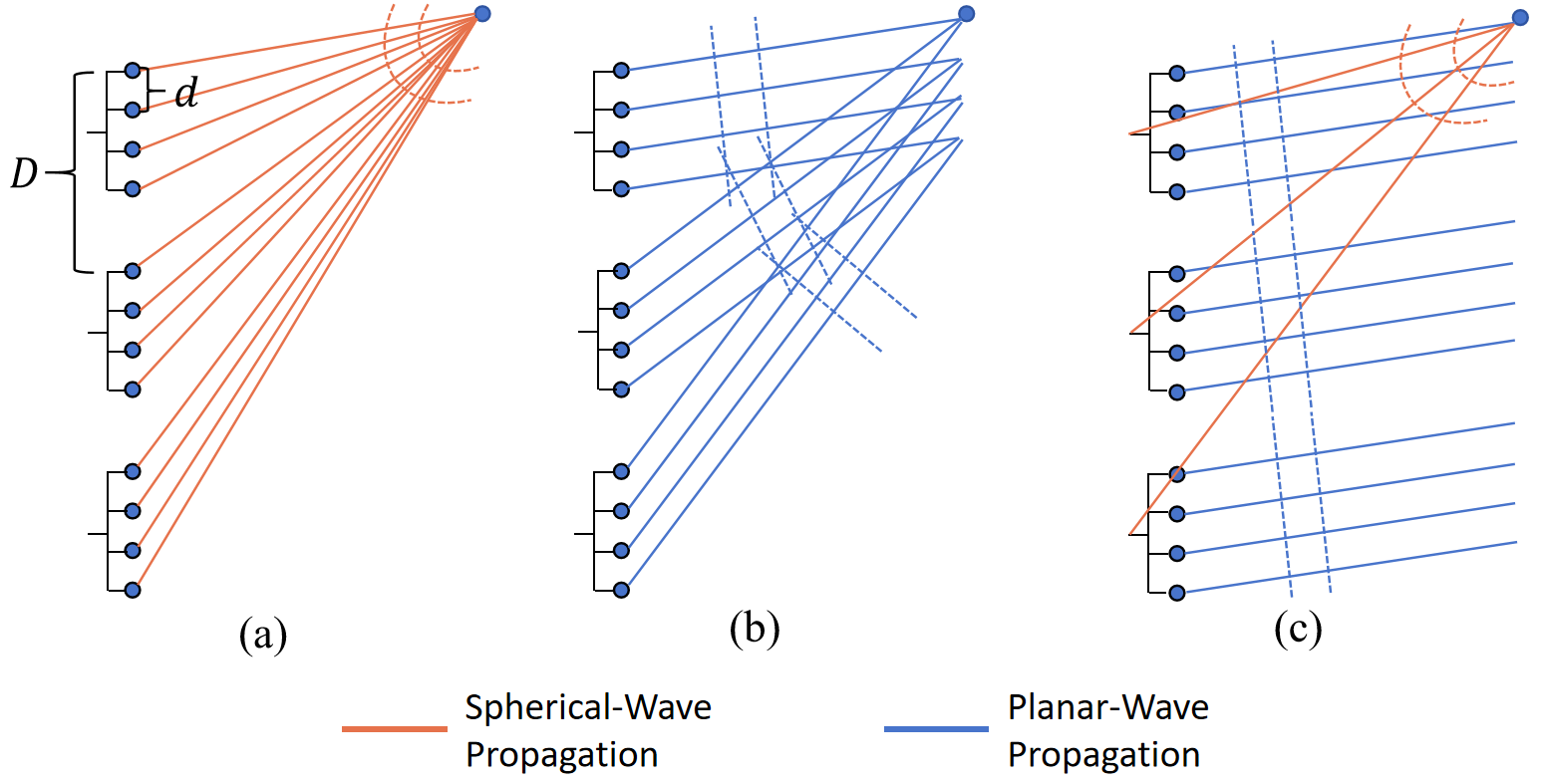}
	\caption{ Visualization of different propagation models: (a) the spherical-wave propagation, (b) planar-wavefront propagation for each subarray, and (c) antenna-level planar-wave and subarray-level spherical-wave propagation.}
	\label{figure1}
\end{figure*}

 Conisder the cross-field model for WSMSs with the HSPW assumption, Eqn. (\ref{g}) is approximated by 
 \begin{equation}\label{cross}
 \mathbf{g}(\theta,r)\approx \mathbf{b}(\theta,r)\otimes\mathbf{a}(\theta),
 \end{equation}
where $\mathbf{a}(\theta)\in\mathbb{C}^{N\times 1}$ denotes the PW array manifold within each subarray and $\mathbf{b}(\theta,r)\in\mathbb{C}^{M\times 1}$ denotes the spherical-wave array manifold for the inter-subarray. They are given by 
\begin{equation}
\mathbf{a}(\theta)\triangleq\frac{1}{N}\left[1 ,\cdots,e^{j\frac{2\pi}{\lambda} d\sin(\theta)},\cdots,e^{j\frac{2\pi}{\lambda} Nd\sin(\theta)} \right]^T,
\end{equation}
\begin{equation}
	\mathbf{b}(\theta,r)\triangleq\frac{1}{M}	\left[ e^{-j\frac{2\pi}{\lambda} (\widetilde{r}^{(1)}-r)},\cdots,e^{-j\frac{2\pi}{\lambda} (\widetilde{r}^{(m)}-r)},\cdots,e^{-j\frac{2\pi}{\lambda} (\widetilde{r}^{(M)}-r)} \right]^T,
\end{equation}
where $\widetilde{r}^{(m)}\triangleq \sqrt{r^2-2r(m-1)D\sin(\theta)+((m-1)D)^2}$, $m\in\{1,\cdots,M\}$.
Indeed, Eqn. (\ref{cross}) can be further approximated by 
 \begin{equation}\label{cross2}
 	\begin{aligned}
 		\mathbf{g}(\theta,r) &\approx \mathbf{b}(\theta,r)\otimes\mathbf{a}(\theta) \\ &= \left[\mathbf{a}(\theta), [\mathbf{b}(\theta,r)]_2\mathbf{a}(\theta),\cdots, [\mathbf{b}(\theta,r)]_M \mathbf{a}(\theta) \right]^T \\
 		& \approx \left[ \mathbf{a}(\theta), \mathbf{a}(\theta_2),\cdots, \mathbf{a}(\theta_M) \right]^T,
 	\end{aligned}
\end{equation}
where $\theta_m, m=2,\cdots,M$ represents the angle of the $m$-th subarray, which is spatially correlated to $\theta$, as illustrated in Fig. \ref{figure1}(b).
 \subsection{Dcitionary Representation for Near-Field Channels}\label{DR}
 For sparse and parametric channel estimation, a key aspect is to represent the channel using a dictionary. This enables the recovery of channel parameters for subsequent channel reconstruction.
 Planar-Wave channels are commonly approximated using a Discrete Fourier Transform (DFT) dictionary, which can orthogonalize the angular-domain beamspace, especially when $d=\lambda /2$.
  We show in Fig. \ref{figure1} three propagation ways for WSMSs. 
In Fig. \ref{figure1}(a), corresponding to Eqn. (\ref{g}), all elements adhere to the spherical-wave propagation pattern.
In Fig. \ref{figure1}(b), corresponding to Eqn. (\ref{cross2}),
 each subarray behaves as a far-field point, steering at a different angle. This corresponds to the separate estimation approach, which individually estimates the parameters of each subarray. In Fig. \ref{figure1}(c), corresponding to Eqn. (\ref{cross}), each subarray steers at the same angle (determined by the reference point), but its weighting is influenced by the subarray-level spherical-wave array manifold.
 	Based on the three models, we propose three distinct virtual representations of near-field channels, each offering unique opportunities for estimation algorithm design. Furthermore, it is worth noting that Fig. \ref{figure1}(b) presents a separate array processing model for each subarray, making it suitable for distributed XL-arrays when we consider the $M$ subarrays as $M$ widely-spaced distributed arrays. Additionally, the model depicted in Fig. \ref{figure1}(c) facilitates distributed signal processing by solely requiring low-dimensional spherical-wave coefficients, effectively reducing centralized processing complexity. 
 \subsubsection{PD-R}
 
 This representation relies on the spherical-wave model for the whole array, i.e., Fig. \ref{figure1}(a).
 For the spherical-wave channel, the beamspace necessitates simultaneous sampling of both angle and distance. This joint sampling can be implemented through uniform sampling of angle-distance pairs.
 	The orthogonality of the dictionary significantly influences the recovery performance, and achieving orthogonality is crucial to using a small number of atoms to characterize the entire angle-distance beamspace. In \cite{NCE1}, a PD dictionary was introduced by the authors to create an approximately orthogonal angle-distance dictionary tailored for spherical-wave channels. The authors of \cite{NCE1} demonstrated that the PD dictionary $\mathbf{G}$ can maintain approximate orthogonality, particularly when the number of distance samples is low. 
 	Essentially, the generation of the PD dictionary follows the principles of the DFT matrix for angle sampling, combined with uniform sampling of distance reciprocals. In the WSMS structure, the PD dictionary proves to be effective in the angle-distance sampling method as well. In the following, we establish a general angle-distance representation $\mathbf{f}_{\rm PDR}$ for near-field WSMS channels:
 \begin{equation}\label{h1}
 \widetilde{\mathbf{h}}\approx\mathbf{f}_{\rm PDR}\triangleq\mathbf{G}\bm{\xi},
 \end{equation}
where $\mathbf{G}\triangleq [\mathbf{g}(\theta_1,r_1),\cdots,\mathbf{g}(\theta_A,r_1),\mathbf{g}(\theta_1,r_2),\cdots,\mathbf{g}(\theta_A,r_C)]\in\mathbb{C}^{MN\times G}$
 represents the PD dictionary, comprising $G \triangleq AC$ atoms. Each atom denotes one angle-distance sample, and
$\bm{\xi}\in\mathbb{C}^{G\times 1}$ is a sparse signal where the non-zero element value signifies the channel path gain, while the non-zero element index indicates the atoms involved in shaping the channel.

  \subsubsection{MAD-R}
As illustrated in Fig. \ref{figure1}(b), different subarrays within a planar wavefront align with a direction of angle. This drives the angular-domain dictionary representation for each array respectively. i.e.,
  \begin{equation}\label{h2}
  	\widetilde{\mathbf{h}}\approx\mathbf{f}_{\rm MADR}\triangleq \frac{1}{\sqrt{P}} \begin{bmatrix}
  	\mathbf{A}\bm{\eta}_1 \\\mathbf{A}\bm{\eta}_2\\ \vdots \\ \mathbf{A}\bm{\eta}_M
  	\end{bmatrix},
  \end{equation}
where $\mathbf{A} \triangleq [\mathbf{a}(\theta_1),\cdots,\mathbf{a}(\theta_A)]\in\mathbb{C}^{N\times A}$ represents the angle dictionary, comprising $A$ atoms in which each atom denotes one angle sample. The sampling angles $\{\theta_1,\cdots,\theta_A\}$ can take values from the DFT bin, specifically $\sin\theta_a\in\{-1,\cdots,-1+\frac{2a}{A},\cdots,1-\frac{2}{A}\}$. Additionally,
$\bm{\eta}_m\in\mathbb{C}^{N\times 1}$, $m\in\{1,\cdots,M\}$, is the sparse signal for the $m$-the subarray in which each non-zeron entry indicates one angle and one path gain for the $m$-th subarray. 
 \subsubsection{2D-PAD-R}
Noticing the array manifold $\mathbf{g}(\theta,r)$ can be approximated by the Kronecker product, this also shows in Fig. \ref{figure1}(c), it can be reshaped into an matrix with a closed-form expression, i.e.,  
 \begin{equation}\label{G}
{\rm devec}(\mathbf{g}(\theta,r))\approx\mathbf{T}(\theta,r)\triangleq \mathbf{a}(\theta) \mathbf{b}^T(\theta,r),
 \end{equation}
where ${\rm devec}(\cdot)$ denotes the devectorization operator. 

Moreover, the subarrary channels $\{\mathbf{h}_m\}_{m=1}^M$ can be stacked in an matrix form of $\mathbf{H}=[\mathbf{h}_1,\cdots,\mathbf{h}_M]\in\mathbb{C}^{N\times M}$. Combining Eqs. (\ref{G}) and (\ref{H}), we have 
\begin{equation}\label{ab}
	\begin{aligned}
		{\rm devec}\left (\widetilde{\mathbf{h}}\right)&=\mathbf{H} \\ & \approx\sqrt{\frac{MN}{L}}\sum_{l=1}^{L} z_l \mathbf{T}(\theta_l,r_l)\\ &=\sqrt{\frac{MN}{L}}\sum_{l=1}^{L} z_l \mathbf{a}(\theta_l) \mathbf{b}^T(\theta_l,r_l).
	\end{aligned}
\end{equation}

In this linear matrix representation, we can employ a 2D atom form to sparsely characterize the coefficient matrix. However, it's important to note that in Eqn. (\ref{ab}), both the row space $\mathbf{a}(\theta)$ and the column space $\mathbf{b}(\theta,r)$ are dependent on the angle parameter $\theta$. This implies that it becomes challenging to create a closed-form 2D sparse representation for it. To address this, the non-zero entry associated with the angle parameter needs to be constrained to maintain a consistent angle for both the row and column spaces. Consequently, $\mathbf{H}$ can be sparsely represented with the constraint:
\begin{equation}\label{h3}
\mathbf{H}=\mathbf{A}\bm{\Xi}\mathbf{B}^T, {\rm s.t.}\ \text{Non-zeros of $\bm{\Xi}$ only on block diagonals},
\end{equation}
where $\mathbf{A}\in\mathbb{C}^{N\times A}$ represents the angle dictionary comprising $A$ atoms,  $\mathbf{B}\in\mathbb{C}^{M\times B}$ denotes the angle-distance dictionary containing $B=A\times C$ atoms, $C$ represents the number of distance samples, and $A$ denotes the number of angle samples. In this context, $\mathbf{A}$ can be defined in the same manner as below Eqn. (\ref{h2}), and $\mathbf{B}$ can be defined similarly to $\mathbf{G}$, with the antenna dimension adjusted to $M$. Morevoer, $\bm{\Xi}$ is a sparse matrix, in which non-zero elements only exist on block diagonals.
 
 Despite this sparse representation, its utilization in recovery algorithms remains challenging due to its NP-hard constraint. Furthermore, to eliminate the need for redundant angle sampling for both the row and column spaces, we design a 2D dictionary that simultaneously samples both angle and distance parameters to address this issue. Consequently, the devectorized channel can be expressed as
\begin{equation}\label{h4}
	\mathbf{H}=\sum_{a\in\{1,\cdots,A\}}\sum_{c\in\{1,\cdots,C\}} \kappa_{a,c}\bm{\mathcal{D}}_{a,c},
\end{equation}
 where $\bm{\mathcal{D}}\in\mathbb{C}^{N\times M\times A\times C}$ is a fourth-order tensor, which we term it as the 2D PAD dictionary, $\{\{\kappa_{a,c}\}\}_{a=1,c=1}^{A,C}$ are the sparse path gains such that $\kappa_{a,c}\in\{0,\kappa_{a,c}\}$, and the non-zero value holds when the 2D PAD atom $\bm{\mathcal{D}}_{a,c}$ is selected. $\bm{\kappa}$ is used to store $\kappa_{a,c}$, for $\forall a,c$, such that $\bm{\kappa}$ is a sparse vector.
  Moreover, $\bm{\mathcal{D}}_{a,c}$ is formed as $\bm{\mathcal{D}}_{a,c}=\mathbf{a}(\theta_a)\mathbf{b}^T(\theta_a,r_c)$, where $\theta_a$ and $r_c$ denote the $a$-th angle sample and the $c$-th distance sample of the 2D PAD dictionary. It is observed from Eqn. (\ref{ab}) that this representation can be well applied for $\mathbf{H}$
  due to $\mathbf{T}(\theta_a,r_c)=\bm{\mathcal{D}}_{a,c}$.
 
  \section{Estimation Frameworks for Near-Field WSMSs}\label{EF}
According to the four sparse representation ways for near-field channels, as shown in Eqs. (\ref{h1}), (\ref{h2}), (\ref{h3}), (\ref{h4}), we propose four estimation frameworks for near-field WSMS channel estimation, which named PD-E, MAD-E, TS-PAD-E, 2D-PAD-E, respectively.
  
	  \subsection{PD-E Framework}\label{DEF}
	  This framework adheres to the conventional approach for near-field channel estimation, directly estimating both angle and distance parameters using the PD dictionary $\mathbf{G}$. Leveraging the signal collected in Eqn. (\ref{y_tilde}) and applying Eqn. (\ref{h1}), the following problem is formulated:
  \begin{equation}\label{DE}
  \begin{aligned}
  & \underset{\bm{\xi}}{\rm arg \ min} \left\Vert \bm{\xi} \right\Vert_0 \\
  & {\rm s.t.} \ \left\Vert\widetilde{\mathbf{y}}-\widetilde{\mathbf{W}}^H\mathbf{G}\bm{\xi}\right\Vert_2^2\leq \epsilon, 
  \end{aligned}
 \tag{$\mathcal{P}_1$}
  \end{equation}
where , and $\epsilon$ is the precise factor. However, this framework requires expensive computational complexity due to the large-scale dictionary such that it is not applicable for practical XL-systems.
  
    \subsection{MAD-E Framework}\label{SEF}
    Examining Eqn. (\ref{h2}), it becomes apparent that the entire channel $\widetilde{\mathbf{h}}$
    can be reconstructed using the set $\{\bm{\eta}_m\}_{m=1}^M$ 
    with the dictionary $\mathbf{A}$. For each subarray, an independent angle recovery problem is formulated for $\forall m$: 
    \begin{equation}\label{SE}
    	\begin{aligned}
    		& \underset{ \bm{\eta}_m}{\rm arg \ min}\ \left\Vert \bm{\eta}_m\right\Vert_0 \\
    		& {\rm s.t.} \ \left\Vert{\mathbf{y}}_m-\mathbf{W}^H\mathbf{A}\bm{\eta}_m\right\Vert_2^2\leq \epsilon.
    	\end{aligned}
    	\tag{$\mathcal{P}_2$}
    \end{equation}

  This framework can effectively reduce the complexity of problem (\ref{DE}) with a smaller dictionary. However, its limitation lies in its neglect of distance estimation and the underutilization of spatial correlation. This may lead to a poor estimation performance.
    \subsection{TS-PAD-E Framework}\label{UCEF}
      Recalling Eqn. (\ref{y_tilde}) and utilizing Eqn. (\ref{cross}), the signal model is re-written by
    \begin{equation}
    	\begin{aligned}
    		\widetilde{\mathbf{y}}=& \left(\mathbf{I}_M\otimes\mathbf{W}^H\right) {\rm vec}(\mathbf{H}).
    	\end{aligned}
    \end{equation}
    
    With the devectorization for $\widetilde{\mathbf{y}}$, we obtain
    \begin{equation}\label{YWH}
    	\mathbf{Y}\triangleq {\rm devec}(\widetilde{\mathbf{y}})= \mathbf{W}^H \mathbf{H}.
    \end{equation}

Furthermore, combining Eqn. (\ref{h3}) yields
\begin{equation}
	\mathbf{Y}= \mathbf{W}^H \mathbf{A}\bm{\Xi}\mathbf{B}^T.
\end{equation}

Since the sparse matrix follows a constraint that the non-zero elements must exist on the block diagonals, we define $\overline{\bm{\Xi}}\triangleq\bm{\Xi}\mathbf{B}^T$ to seek for a two-stage recovery solution. In this sense, $\overline{\bm{\Xi}}$ is a row-sparse matrix without any other constraint. This directly formulates an MMV problem:
    \begin{equation}\label{MMV}
	\begin{aligned}
		& \underset{\overline{\bm{\Xi}}}{\rm arg \ min}  \left\Vert {\rm diag}\left\{\overline{\bm{\Xi}}\overline{\bm{\Xi}}^H\right\}\right\Vert_0 \\
		& {\rm s.t.} \   \left\Vert{\mathbf{Y}}-\mathbf{W}^H\mathbf{A}\overline{\bm{\Xi}}\right\Vert_2^2\leq \epsilon,
	\end{aligned}
	\tag{$\mathcal{P}_3$}
\end{equation}
where ${\rm diag}\{\cdot\}$ represents extracting the diagonal elements of a matrix.

After $\overline{\bm{\Xi}}$ is recovered, we denote by $\widehat{\bm{\Xi}}$ the matrix obtained by removing rows with zero elements from $\overline{\bm{\Xi}}$, and the non-zero index set is denoted by ${\Gamma}$ such that $\widehat{\bm{\Xi}}=\left[\overline{\bm{\Xi}}\right]_{\Gamma,:}$.
Moreover, assuming a perfect recovery indicates   $\widehat{\bm{\Xi}}=[{\bm{\Xi}}]_{\Gamma,:}\mathbf{B}^T$. For each row of $\widehat{\bm{\Xi}}$, it should be a $1$-sparse vector. Thus, we can formulate a pallelizable recovery problem to find ${\rm card}(\Gamma)$ nonzeros in those ${\rm card}(\Gamma)$ rows with the dictionary $\mathbf{B}$.

The above two-stage recovery framework can be understood in another way as following.
   Referring to Eqn. (\ref{cross}), we can decouple the $MN$-dimentional vector $\mathbf{g}(\theta,r)$ into $M$ $N$-dimentional vectors $\{[\mathbf{b}(\theta,r)]_1\mathbf{a}(\theta),[\mathbf{b}(\theta,r)]_2\mathbf{a}(\theta),\cdots,[\mathbf{b}(\theta,r)]_M\mathbf{a}(\theta)\}$. Alternatively,  we can express this as the weighting of  $\mathbf{a}(\theta)$ by $\mathbf{b}(\theta,r)$. 
   This encourages the joint recovery of $\mathbf{a}(\theta)$ alongside the signals from the $M$ subarrays. Subsequently, this allows for the recovery of $\mathbf{b}(\theta,r)$ by combining the estimated channel path gain with the weight structure.
   \subsection{2D-PAD-E Framework}\label{2DP}
 The two-stage estimation process offers an ingenious solution to address the sparsity constraint in Eqn. (\ref{h3}). However, one drawback is the potential for error accumulation when estimating the distance parameter. As a result, we delve into further exploration to design a more efficient estimation framework, leveraging the fourth method to represent the near-field channel. 
   
Combining Eqs. (\ref{h4}) and (\ref{YWH}) yields
\begin{equation}
	\mathbf{Y}=\sum_{a\in\{1,\cdots,A\}}\sum_{c\in\{1,\cdots,C\}} \kappa_{a,c}\mathbf{W}^H\bm{\mathcal{D}}_{a,c}.
\end{equation}

In this expression, we aim for $\bm{\kappa}\triangleq\{ \kappa_{a,c}|a=1,\cdots,A, c=1,\cdots,C\}$ to exhibit sparsity. Unlike Eqn. (\ref{h3}), there is no sparsity constraint imposed on the positions of non-zero elements. Therefore, a one-stage solution can be contemplated. Consequently, we formulate a 2D recovery problem that leverages the 2D dictionary as
\begin{equation}\label{2DE}
	\begin{aligned}
		  &\underset{ \bm{\kappa}}{\rm arg \ min}\ \left\Vert \bm{\kappa}\right\Vert_0 \\
		  {\rm s.t.} \ & \left\Vert\mathbf{Y}-\sum_{a\in\{1,\cdots,A\}}\sum_{c\in\{1,\cdots,C\}} \kappa_{a,c}\mathbf{W}^H\bm{\mathcal{D}}_{a,c}\right\Vert_F^2\leq \epsilon.
	\end{aligned}
	\tag{$\mathcal{P}_4$}
\end{equation}

As this problem requires calculations with a 2D dictionary, its recovery process deviates from the standard methods outlined in problems (\ref{DE}) and (\ref{SE}). In Section \ref{2D-PAD-OMP}, we present a greedy solution that incorporates 2D PAD atom selection to tackle this unique challenge.

In summary, the PD-E framework necessitates all antennas to collectively process for angle-distance estimation, resulting in high complexity and rendering it unsuitable for decentralized systems. Conversely, MAD-E completely circumvents reliance on joint design by allowing each subarray to independently estimate its own angle. However, this approach may incur significant performance loss. As a balanced design approach, both TS-PAD-E and 2D-PAD-E adopt a subarray-level joint processing strategy. The former employs a two-stage angle-distance decoupled estimation, while the latter performs joint angle-distance estimation. Both TS-PAD-E and 2D-PAD-E can be tailored to facilitate distributed XL-array systems, as elaborated in Section \ref{DR}.

  \begin{algorithm}
	\caption{Procedure of TS-PAD-OMP} 
	\label{AL1}
	\KwIn {Measured signals $\mathbf{Y}$, measurement matrix $\mathbf{W}$, dictionary $\mathbf{A}$, sparsity $L$, $\Lambda=\emptyset$, and $\mathcal{A}$ representing the atom index set.
	}
	\KwOut {The estimated channel $\widehat{\widetilde{\mathbf{h}}}$.}
	\Begin{ 
		\emph{\%\% Angle Estimation} 
		\\
		$\mathbf{R}^{(0)}=\mathbf{Y}$
		\\
		\For{$l=1,\cdots,L$}{ 
			$ a^\star=	 \underset{a}{\rm arg\ max } \ \sum_{m=1}^M \frac{ \left\vert[\mathbf{A}]_{:,a}^H\mathbf{W}[\mathbf{R}^{(l-1)}]_{:,m}\right\vert}{\left\Vert [\mathbf{A}]_{:,a}^H\mathbf{W}\right\Vert_2^2}$
			\\
			$\Lambda\leftarrow\Lambda\cup a^\star$, $\mathcal{A}\leftarrow\mathcal{A}\setminus a^\star$\\
			$\bm{\Phi}^{(l)}\triangleq \mathbf{W}^H[\mathbf{A}]_{:,
				\Lambda}$\\
			$\mathbf{R}^{(l)}=\bm{\Phi}^{(l),\perp}\mathbf{R}^{(l-1)}$
		} 	Obtain $\widehat{\bm{\Xi}}=\left(\bm{\Phi}^{(l),H}\bm{\Phi}^{(l)}\right)^{-1}\bm{\Phi}^{(l),H}\mathbf{Y}$ \\	\emph{\%\% Distance Estimation}  \\
		\For{$l=1,\cdots,L$}{
			$\widehat{\mathbf{z}}_l=[\widehat{\bm{\Xi}}]_{l,:}^T$ \\
			$\widehat{r}_l=\underset{r_l}{\rm arg \ max} \    {\left\vert  \mathbf{b}^H(\widehat{\theta}_l,r_l)\widehat{\mathbf{z}}_l\right\vert}$
		}
		\emph{\%\% Channel Reconstruction} \\
		$\widehat{\mathbf{G}}=\left[
		{\mathbf{g}}\left(\widehat{\theta_1},\widehat{r}_1\right),\cdots,{\mathbf{g}}\left(\widehat{\theta}_L,\widehat{r}_L\right)\right]$\\
		$\widehat{\widetilde{\mathbf{h}}}=\widehat{\mathbf{G}}\left(\widetilde{\mathbf{W}}^H\widehat{\mathbf{G}}\right)^\dagger\widetilde{\mathbf{y}}$
	}
	\Return{$\widehat{\widetilde{\mathbf{h}}}$}
\end{algorithm}

 \section{Estimation Procedures Based on Orthogonal Matching Pursuit}\label{EOMP}
 OMP is a classical greedy algorithm, which can be used for sparse signal recovery in a simple implementation and with low complexity. Therefore, it is very suitable for us to evaluate the four estimation frameworks in the last section. By combining the four framworks with OMP, we term them as PD-OMP, MAD-OMP, TS-PAD-OMP, and 2D-PAD-OMP, respectively. Since PD-OMP and MAD-OMP can be simply solved by OMP with the given dictionary, we focus on discussing TS-PAD-OMP and 2D-PAD-OMP as follows.   
 Beyond the impact of sparse recovery methods on estimation effectiveness, the measurement matrix described in Eqn. (\ref{y_tilde}) plays a pivotal role in estimation performance. Thus, we investigate the optimization of the measurement matrix to enhance estimation accuracy. Furthermore, we conduct an analysis of the computational complexity associated with the four proposed methods.
 \subsection{TS-PAD-OMP}
 Here, we solve the near-field channel estimation problem by the proposed TS-PAD-E framework based on OMP. It includes two stages: joint angle estimation and distance estimation. 
 
 \begin{algorithm}
 	\caption{Procedure of 2D-PAD-OMP} 
 	\label{AL2}
 	\KwIn {Measured signals $\mathbf{Y}$, measurement matrix $\mathbf{W}$, dictionary $\bm{\mathcal{D}}$, sparsity $L$, $\Gamma=\emptyset$, and $\mathcal{A}$ representing the atom index set.
 	}
 	\KwOut {The estimated channel $\widehat{\widetilde{\mathbf{h}}}$.}
 	\Begin{  
 		$\mathbf{R}^{(0)}=\mathbf{Y}$
 		\\
 		\For{$l=1,\cdots,L$}{ 
 			$ 	(a^\star,c^\star)=	\underset{a,c}{\rm arg\ max } \  \frac{\left\vert\left\langle\mathbf{W}\mathbf{R}^{(l)},\bm{\mathcal{D}}_{a,c}\right\rangle\right\vert}{\left\Vert \mathbf{W}^H[\mathbf{A}]_{:,a}\right\Vert_2^2}$
 			\\
 			$\Gamma\leftarrow\Gamma\cup (a^\star,c^\star)$, $\mathcal{I}\leftarrow\mathcal{I}\setminus (a^\star,c^\star)$\\
 			$\mathbf{X}\leftarrow\left[\mathbf{X},{\rm vec}\left(\mathbf{W}^H\bm{\mathcal{D}}_{a^\star,c^\star}\right)\right]$\\
 			$	\bm{\kappa}_{\Gamma}=\left(\mathbf{X}^H\mathbf{X}\right)^{-1}\mathbf{X}^H{\rm vec}(\mathbf{R}^{(l-1)})$
 			\\
 			$  \mathbf{R}^{(l)}=\mathbf{R}^{(l-1)}-\sum_{(a,c)\in\Gamma}  \kappa_{a,c}\mathbf{W}^H\bm{\mathcal{D}}_{a,c}$ 
 		} 
 		\emph{\%\% Channel Reconstruction} \\
 		$\bm{\kappa}_{\Gamma}=\left(\mathbf{X}^H\mathbf{X}\right)^{-1}\mathbf{X}^H\widetilde{\mathbf{y}}$\\	$\widehat{\mathbf{H}}=\sum_{(a,c)\in\Gamma}  \kappa_{a,c} \bm{\mathcal{D}}_{a,c}$ \\  
 		$\widehat{\widetilde{\mathbf{h}}}={\rm vec}\left(\widehat{\mathbf{H}}\right)$
 	}
 	\Return{$\widehat{\widetilde{\mathbf{h}}}$}
 \end{algorithm}
 \subsubsection{Stage 1: Joint Angle Estimation}\label{MMV-OMP}
 The TS-PAD-E framework begins with an MMV problem for $\overline{\bm{\Xi}}$ recovery, i.e., solving
 problem (\ref{MMV}). For this MMV problem, SOMP can be well adopted. It follows two main steps: 1) atom identifying with $\ell_1$ norm, and 2) residual calculation via least squares (LS).
 
 \emph{Atom Identifying:} The criterio of SOMP to select the atom in the $l$-th iteration is  
 \begin{equation}
 \underset{a}{\rm arg\ max } \ \sum_{m=1}^M \frac{ \left\vert[\mathbf{A}]_{:,a}^H\mathbf{W}[\mathbf{R}^{(l)}]_{:,m}\right\vert}{\left\Vert [\mathbf{A}]_{:,a}^H\mathbf{W}\right\Vert_2^2}
 \end{equation} 
 where $\mathbf{R}^{(l-1)}$ is the residual signal of the $(l-1)$-th iteration.
 
  \emph{Residual Calculation:} After selecting one atom, the residual signal of the $l$-th iteration will be updated by
  \begin{equation}
  	\mathbf{R}^{(l)}=\bm{\Phi}^{(l),\perp}\mathbf{R}^{(l-1)},
  \end{equation}  
where $\bm{\Phi}^{(l)}\triangleq \mathbf{W}^H[\mathbf{A}]_{:,\Lambda^{(l)}}$, $\Lambda^{(l)}$ is the selected atom index set in the $l$-th iteration, and $\bm{\Phi}^{(l),\perp}\triangleq \mathbf{A}-\bm{\Phi}^{(l)}\left(\bm{\Phi}^{(l),H}\bm{\Phi}^{(l)}\right)^{-1}\bm{\Phi}^{(l),H}$.
  \subsubsection{Stage 2: Distance Estimation} 
  According to the estimated $\widehat{\bm{\Xi}}$ in the last section, we denote by $\widehat{\mathbf{z}}_l\triangleq\left[\widehat{\bm{\Xi}}\right]_{l,:}^T\in\mathbb{C}^{M\times 1}$ the $l$-th row of $\widehat{\bm{\Xi}}$. 
  Recalling Eqs. (\ref{H}) and (\ref{cross}), we find that
  $\widehat{\mathbf{z}}_l\approx\overline{\mathbf{z}}_l\triangleq\sqrt{\frac{MN}{L}}z_l\left[[\mathbf{b}(\theta_l,r_l)]_1,\cdots,[\mathbf{b}(\theta_l,r_l)]_M\right]^T=\sqrt{\frac{MN}{L}}z_l\mathbf{b}(\theta_l,r_l)$.
  Thus, the distance recovery problem can be formulated by
  \begin{equation} \underset{r_l}{\rm arg \ max} \    {\left\vert  \mathbf{b}^H(\widehat{\theta}_l,r_l)\widehat{\mathbf{z}}_l\right\vert},
  \end{equation}
  where $\left\{\widehat{\theta}_l\right\}_{l=1}^L$ are the estimated angles in last stage.
  
  In this manner, $\{\widehat{r}_l\}_{l=1}^L$ can be estimated. 
  Finally, the channel support is formed given by $\widehat{\mathbf{G}}=\left[
  {\mathbf{g}}\left(\widehat{\theta_1},\widehat{r}_1\right),\cdots,{\mathbf{g}}\left(\widehat{\theta}_L,\widehat{r}_L\right)\right]$. Thereby
  the WSMS channel is reconstructed by $\widehat{\widetilde{\mathbf{h}}}=\widehat{\mathbf{G}}\left(\widetilde{\mathbf{W}}^H\widehat{\mathbf{G}}\right)^\dagger\widetilde{\mathbf{y}}$. The whole procedure is shown in Algorithm \ref{AL1}.
 \subsection{2D-PAD-OMP}\label{2D-PAD-OMP}
Unlike the atom selection criteria of conventional OMP, 2D-PAD-OMP selects the optimal atom based on the 2D projection of the residual signal. The specific details are as follows.
     
    \emph{Atom Identifying:} The criterio of 2D-PAD-OMP to select the atom in the $l$-th iteration is  
  \begin{equation}\label{WR}
	\underset{a,c}{\rm arg\ max } \  \frac{ \left\vert\mathbf{a}^H(\theta_a)\mathbf{W}\mathbf{R}^{(l)}\mathbf{b}^*(\theta_a,r_c)\right\vert}{\left\Vert \mathbf{a}^H(\theta_a)\mathbf{W}\right\Vert_2^2}.
  \end{equation}
For clarity, we denote by $\left\vert\left\langle\mathbf{W}\mathbf{R}^{(l)},\bm{\mathcal{D}}_{a,c}\right\rangle\right\vert\triangleq  \left\vert\mathbf{a}^H(\theta_a)\mathbf{W}\mathbf{R}^{(l)}\mathbf{b}^*(\theta_a,r_c)\right\vert$. Eqn. (\ref{WR}) is equivalent to 
  \begin{equation}\label{WR2}
	\underset{a,c}{\rm arg\ max } \  \frac{\left\vert\left\langle\mathbf{W}\mathbf{R}^{(l)},\bm{\mathcal{D}}_{a,c}\right\rangle\right\vert}{\left\Vert \mathbf{W}^H[\mathbf{A}]_{:,a}\right\Vert_2^2}.
\end{equation}

  \emph{Residual Calculation:} In contrast to the residual calculation method used in conventional OMP, we need to solve a matrix-form coefficient calculation. Given the selected atoms $\bm{\mathcal{D}}_{a,c}$, $(a,c)\in\Gamma$
  for $(a,c)\in\Gamma$, along with the residual signal from the $(l-1)$-th iteration, we obtain 
  \begin{equation}  
  	\begin{aligned} \underset{\bm{\kappa}_\Gamma}{\rm arg \ min}\left\Vert\mathbf{R}^{(l-1)}-\sum_{(a,c)\in\Gamma}  \kappa_{a,c}\mathbf{W}^H\bm{\mathcal{D}}_{a,c}\right\Vert_F^2,
  	\end{aligned}
  \end{equation}
where $\bm{\kappa}_\Gamma$ denotes the ${\rm card}(\Gamma)$ coefficients corresponding to the selected atoms. 

We simplify this problem via the vectorization operator. First, we have
\begin{equation}
	\begin{aligned}
		{\rm vec}\left(\sum_{(a,c)\in\Gamma}  \kappa_{a,c}\mathbf{W}^H\bm{\mathcal{D}}_{a,c}\right)=&\sum_{(a,c)\in\Gamma}  \kappa_{a,c}{\rm vec}\left(\mathbf{W}^H\bm{\mathcal{D}}_{a,c}\right)\\ =& \mathbf{X} {\bm{\kappa}}_\Gamma,
	\end{aligned}
\end{equation} 
where $\mathbf{X}\in\mathbb{C}^{MQ\times {\rm card}(\Gamma)}$ is constructed by stacking $\{{\rm vec}\left(\mathbf{W}^H\bm{\mathcal{D}}_{a,c}\right)|(a,c)\in\Gamma\}$ in columns. 

Then, the solution of ${\bm{\kappa}}_\Gamma$ regarding ${\rm min}\ \Vert{\rm vec}(\mathbf{R}^{(l)})-\mathbf{X} {\bm{\kappa}}_\Gamma\Vert_2^2$ can be obtained by the LS algorithm as 
\begin{equation}
	\bm{\kappa}_{\Gamma}=\left(\mathbf{X}^H\mathbf{X}\right)^{-1}\mathbf{X}^H{\rm vec}(\mathbf{R}^{(l-1)}).
\end{equation}
 
 After $	\bm{\kappa}_{\Gamma}$ is attained, the residual signal can be calculated by
 \begin{equation}
  \mathbf{R}^{(l)}=\mathbf{R}^{(l-1)}-\sum_{(a,c)\in\Gamma}  \kappa_{a,c}\mathbf{W}^H\bm{\mathcal{D}}_{a,c}.
 \end{equation}

 Once the index support $\Gamma$ is obtained, and the coefficients are calculated, the channel can be reconstructed by  $\widehat{\mathbf{H}}=\sum_{(a,c)\in\Gamma}  \kappa_{a,c} \bm{\mathcal{D}}_{a,c}$  and 
 $\widehat{\widetilde{\mathbf{h}}}={\rm vec}\left(\widehat{\mathbf{H}}\right)$. The whole procedure is shown in Algorithm \ref{AL2}.
\subsection{Measurement Matrix Optimization}\label{MMO}

Without any priori information before channel estimation, one possible way to design the measurement matrix is to reduce its total coherence \cite{MMO}:
 \begin{equation}\label{SMO1}
	\begin{aligned}	\underset{\mathbf{W}}{\rm arg \ min} \left\Vert\mathbf{I}_{Q}- {\mathbf{W}}^H{\mathbf{W}}\right\Vert_F^2. 
	\end{aligned}
	\tag{$\mathcal{P}_{5}$}
\end{equation}

Using the SVD of $\mathbf{W}={\mathbf{U}}{\bm{\Sigma}}{\mathbf{V}}^H$ with unitary matrices ${\mathbf{U}}\in\mathbb{C}^{N\times N}$, ${\mathbf{V}}\in\mathbb{C}^{Q\times Q}$ and $\bm{\Sigma}=\left[{\rm diag}(\bm{\sigma}), \mathbf{0}_{Q,N-Q}\right]^T$, where $\bm{\sigma}=[\sigma_1,\cdots,\sigma_{Q}]$ and $\mathbf{0}_{Q,N-Q}\in\mathbb{C}^{Q\times(N-Q)}$ is a null matrix. 
\begin{equation}
	\begin{aligned}
		\left\Vert\mathbf{I}_{Q}-\mathbf{W}^H\mathbf{W}
		\right\Vert_F^2=&\left\Vert \mathbf{I}_{Q}-{\mathbf{U}}{\bm{\Sigma}}{\mathbf{V}}^H{\mathbf{V}}{\bm{\Sigma}}^H{\mathbf{U}}^H
		\right\Vert_F^2\\
		=&\left\Vert {\mathbf{U}} \left(\mathbf{I}_{Q}-{\bm{\Sigma}}{\bm{\Sigma}}^H\right){\mathbf{U}}^H
		\right\Vert_F^2\\
		=&\left\Vert
		\mathbf{I}_{Q}-{\bm{\Sigma}}{\bm{\Sigma}}^H
		\right\Vert_F^2\\
		=&\sum_{i=1}^{Q}(1-\sigma^2_i)^2.
	\end{aligned}
\end{equation}

As $\sigma_1=\cdots=\sigma_Q=1$, the above equation is minimized.  Hence,  $\mathbf{W}= {\mathbf{{U}}}_1[\mathbf{I}_Q, \mathbf{0}_{Q,N-Q}]^T{\mathbf{{V}}}_1^H$ is the unconstraint solution of problem (\ref{SMO1}), where ${\mathbf{{U}}}_1\in\mathbb{C}^{N\times N}$ and ${\mathbf{{V}}}_1\in\mathbb{C}^{Q\times Q}$ are arbitrary unitary matrices. 
For the modulus-1 constraint of phased-arrays, $\mathbf{W}$ is modified as $[\mathbf{W}]_{n,q}=\frac{[\mathbf{W}]_{n,q}}{\vert [\mathbf{W}]_{n,q}\vert}$.
\subsection{Computational Complexity Analysis}\label{CCA}

This section discusses the time complexity for the four estimation methods.  OMP exhibits linear complexity, primarily dominated by the atom identification step, incurring a complexity of $\mathcal{O}(QNL)$ given a $Q$-dimensional measurement signal, a $N\times N$ sensing matrix, and an iteration number $L$. It is worth noting that we assume $G=AC$ for the PD dictionary $\mathbf{G}$. Consequently, the complexity for PD-OMP and MAD-OMP can be expressed as $\mathcal{O}(QMACL)$ and $\mathcal{O}(QMAL)$, respectively.  The distinction arises from the fact that PD-OMP relies on a PD dictionary comprising angle-distance atoms. The complexity of TS-PAD-OMP is predominantly governed by the first MMV estimation stage. Solving the joint angle estimation using SOMP incurs a complexity of $\mathcal{O}(QMAL)$.  Subsequently, the distance estimation in the second stage requires a complexity of $\mathcal{O}(MCL)$. Thus, the total complexity sums up to $\mathcal{O}(QMAL+MCL)$.For 2D-PAD-OMP, the 2D atom identification in Eqn. (\ref{WR2}) can be computed as  $\left[\overline{\mathbf{W}}\right]_{a,:}\mathbf{R}^{(l)}\mathbf{B}^*(\theta_a)$, for $\forall a$, where $\overline{\mathbf{W}}\triangleq\mathbf{A}^H\mathbf{W}\in\mathbb{C}^{A\times Q}$ can be calculated before the online estimation procedure, and
  $\mathbf{B}^*(\theta_a)$ denotes the columns corresponding to $\theta_a$ in $\mathbf{B}$. This operation necessitates a complexity of $\mathcal{O}(QMA+MAC)$. Consequently, the total complexity accumulates to $\mathcal{O}(QMAL+MACL)$. In summary, MAD-OMP exhibits the lowest complexity, while PD-OMP commands the highest. TS-PAD-OMP and 2D-PAD-OMP demonstrate comparable complexities that are substantially lower than that of PD-OMP. 
  
  \begin{figure}
  	\centering
  	\includegraphics[width = 0.49\textwidth]{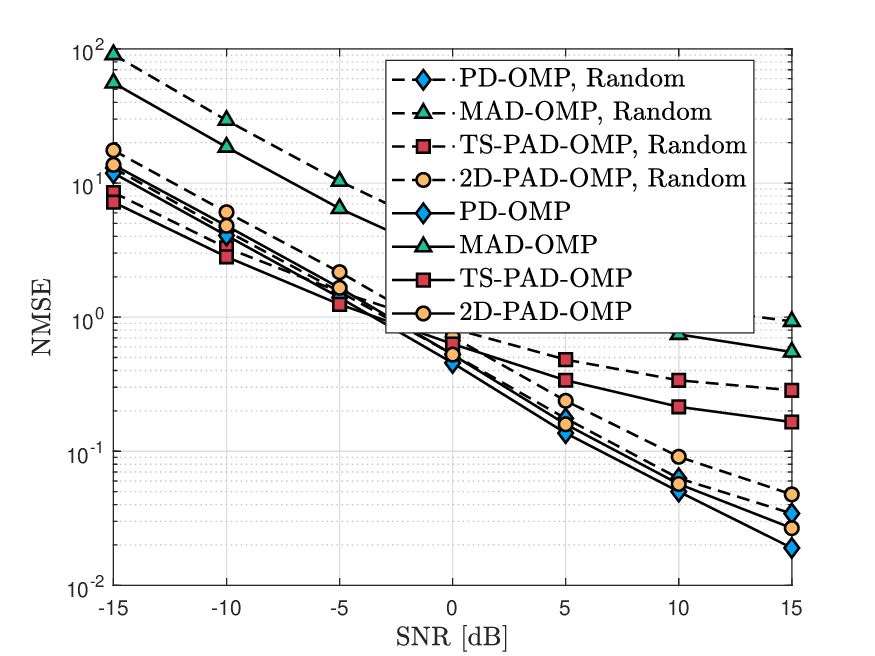}
  	\caption{The NMSE performance of different methods regarding SNR and measurement matrix.}
  	\label{MMOF}
  \end{figure}
  
  \begin{figure}
  	\centering
  	\subfloat[$Q=8$]{
  		\includegraphics[width=2.72in]{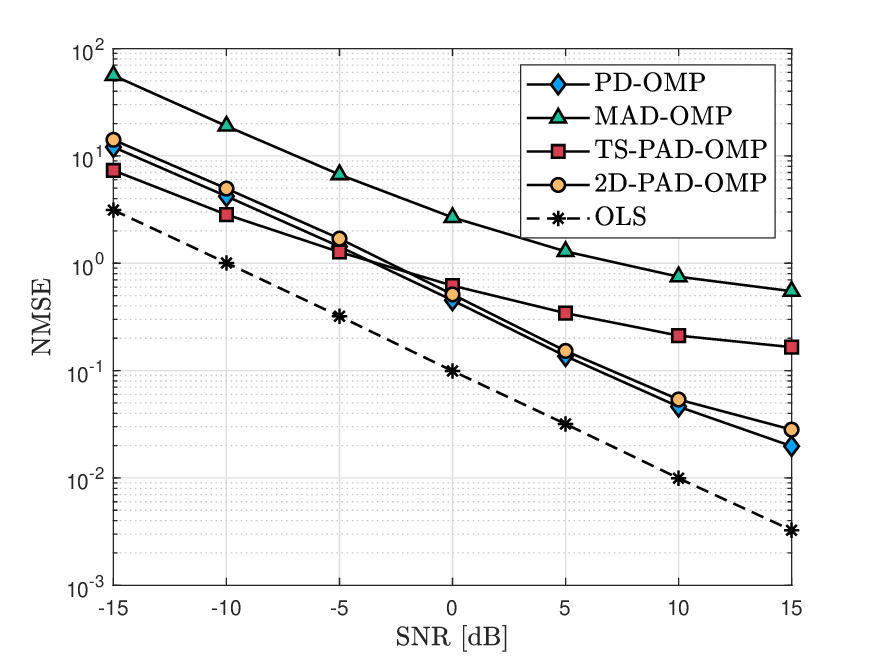}
  	}
  	\quad    
  	\subfloat[$Q=16$]{
  		\includegraphics[width=2.72in]{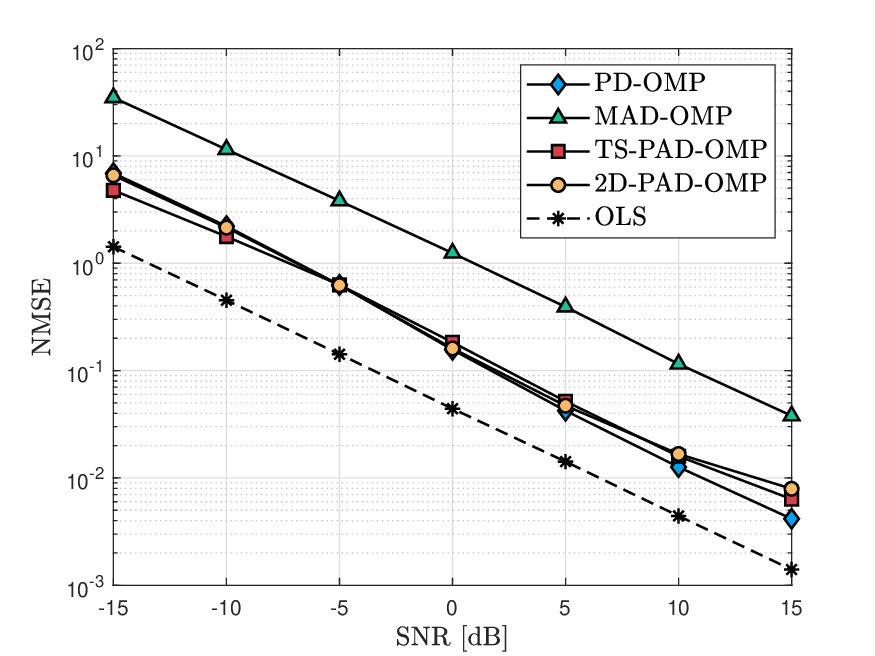}
  	}
  	\caption{The NMSE performance of different methods varies with SNR when $Q=8$ and $Q=16$.}
  	\label{NMSE_SNR}
  \end{figure}
 \section{Simulation Results}\label{SR}
Several numerical simulations are conducted to demonstrate the effectiveness of the proposed method with various system parameters set.
The system adopts $100$ GHz central frequency. The number of RF chains and antennas connect to each RF chain are $M=8$ and $N=24$, respectively. The inter-antenna spacing and inter-subarray spacing are set to $d=\lambda/2$ and $D=Nd+8\lambda$, respectively. The far- and near-field region is bounded by the Fraunhofer distance $R_{\rm NF}=\frac{2(MD)^2}{\lambda}$. 
The user/scatter locations are assumed to be distributed within a sector region where $\sin(\theta)$ ranges from $[-0.75, 0.75]$, distances range from $[5, 2R_{\rm NF}]$ meters, and the grid has a resolution of $2/N$ for the angle and $5$ for the communication distance.
 Since the transmit power is set to 1, the SNR is defined by $\frac{1}{\sigma_n^2}$. 
The benchmarks are described as follows, and all of them adopt the optimized measurement matrix in Section \ref{MMO}:
\begin{itemize}
	\item \textbf{PD-OMP:} Using OMP for the PD-E framework in Section \ref{DEF}.
	\item \textbf{MAD-OMP:} Using OMP for the MAD-E framework in Section \ref{SEF}. 
	\item \textbf{TS-PAD-OMP:} As shown in Algorithm \ref{AL1}.
	\item \textbf{2D-PAD-OMP:} As shown in Algorithm \ref{AL2}.
		\item \textbf{OLS:} Using the oracle least squares (OLS) estimator as the lower bound, which assumes the perfect channel support.
\end{itemize}

In this paper, we choose the normalized mean squared error (NMSE) as the performance metric, calculated by $\mathbb{E}\left\{ \frac{\left\Vert\widetilde{\mathbf{h}}-\widehat{\widetilde{\mathbf{h}}}\right\Vert_2^2 }{\left\Vert \widetilde{\mathbf{h}}\right\Vert_2^2 }\right\}$. We will assess the NMSE of various methods, considering the influence of pilot overhead $Q$, SNR, and the number of channel paths on their performance.

We begin by assessing the impact of measurement matrix optimization on the channel estimation errors of our proposed methods. For this purpose, we plot the NMSE of different methods under the conditions where $L=4$, $Q={8}$, SNR$\in\{-15,-10,-5,0,5,10,15\}$ dB, and the measurement matrix can be optimized as described in Section \ref{MMO}, alongside the "Random" approach, which is generated by a Gaussian random matrix with modulus-1 elements. As illustrated in Fig. \ref{MMOF}, the methods based on random measurement matrices (indicated by dashed lines) exhibit higher NMSE across all methods compared to those based on optimized measurement matrices (indicated by solid lines). This suggests that measurement matrix optimization contributes to improved estimation performance for our proposed methods.
\begin{figure}
	\centering
	\subfloat[SNR $=-5$ dB]{
		\includegraphics[width=2.09in]{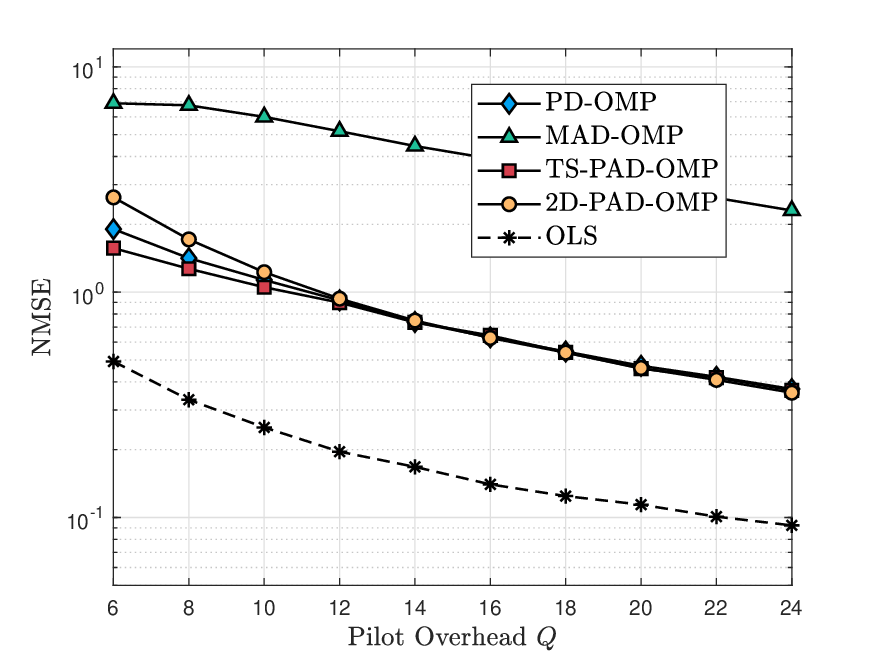}
	}    
	\subfloat[SNR $=5$ dB]{
		\includegraphics[width=2.09in]{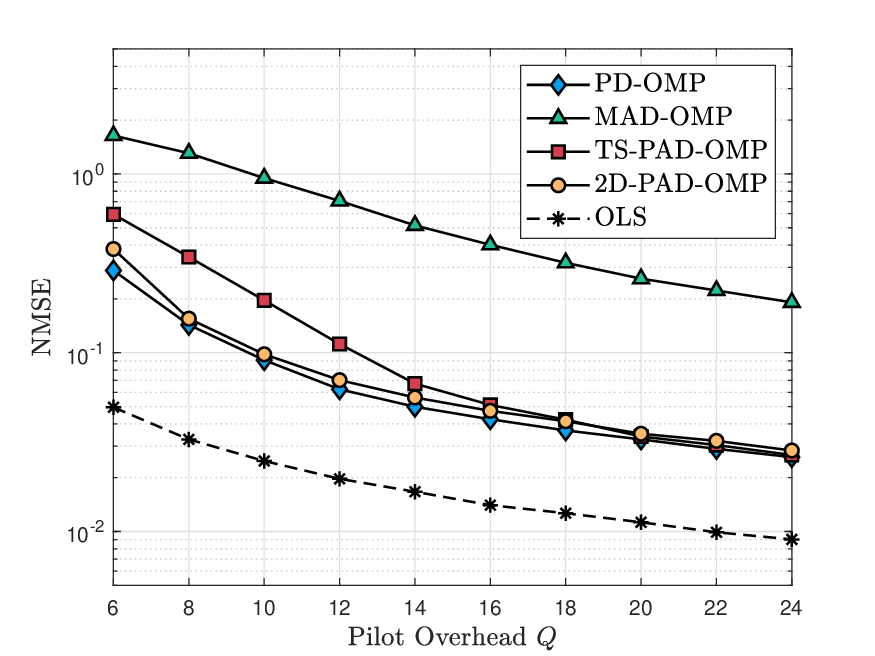}
	}
	\subfloat[SNR $=15$ dB]{
	\includegraphics[width=2.09in]{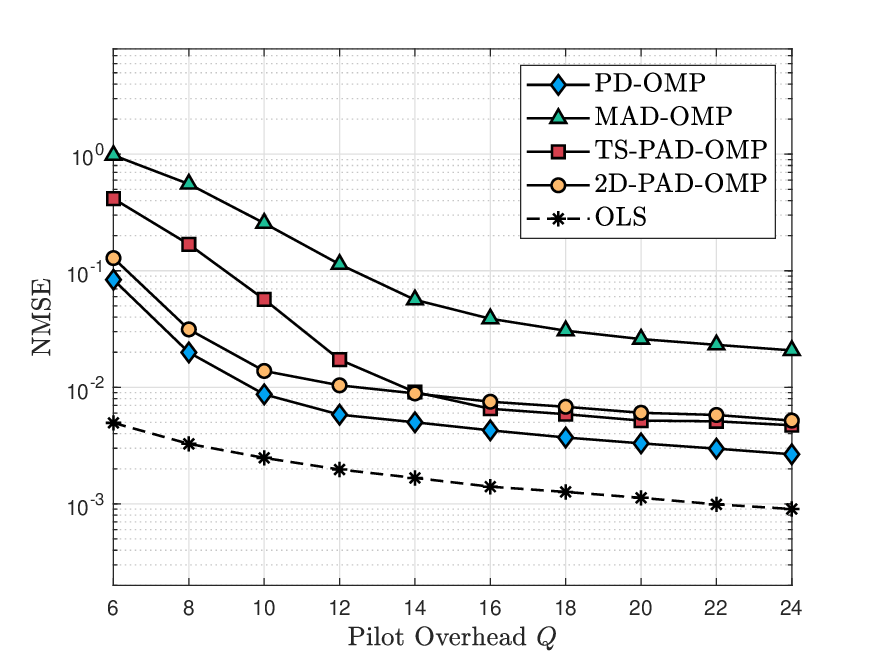}
}
	\caption{The NMSE performance of different methods varies with $Q$ when the SNR is $0$ and $15$ dB.}
	\label{NMSE_Q}
\end{figure}

As shown in Fig. \ref{NMSE_SNR}, we plot the NMSE of different methods in cases when $L=4$, $Q\in\{8,16\}$, and SNR$\in\{-15,-10,-5,0,5,10,15\}$ dB. As observed in Fig. \ref{NMSE_SNR}(a) and from the complexity analysis in Section \ref{CCA}, it is evident that although MAD-OMP has the lowest computational complexity due to its ability to handle multiple parallel small-scale recovery problems, it exhibits the weakest performance among all the methods. Conversely, PD-OMP, despite having the highest complexity, outperforms others owing to the numerous spherical-wave atoms in its dictionary. TS-PAD-OMP demonstrates commendable performance at lower SNR levels when $Q=8$. This could be attributed to the initial stage of TS-PAD-OMP failing to estimate accurate angles at lower $Q$ values in high SNR scenarios, leading to error accumulation in the subsequent stage. Remarkably, 2D-PAD-OMP attains an NMSE performance comparable to that of PD-OMP, highlighting its effectiveness despite its lower complexity. In Fig. \ref{NMSE_SNR}(b), with an increased number of measurements at $Q=16$, it is observed that PD-OMP, TS-PAD-OMP, and 2D-PAD-OMP show similar NMSE performances, nearing the theoretical lower bound. Under these conditions, MAD-OMP also performs satisfactorily. Additionally, a linear relationship between NMSE and SNR is noticeable across the various methods.

In Fig. \ref{NMSE_Q}, we plot the NMSE of different methods in cases when $L=4$, $Q\in\{6,8,\cdots,24\}$, and SNR$\in\{-5,5,15\}$ dB. Overall, the NMSE trends observed are akin to those in Fig. \ref{NMSE_SNR}, with a notable distinction being the non-linear relationship of NMSE with the pilot overhead $Q$. Fig. \ref{NMSE_Q}(a) highlights a small NMSE difference between 2D-PAD-OMP and PD-OMP for $Q\geq8$ at an SNR of $-5$ dB. Conversely, Fig. \ref{NMSE_Q}(c) illustrates a significant NMSE gap between all methods and the theoretical lower bound when $Q$ is low. This suggests that the proposed methods struggle to approach the lower bound with a limited number of measurements, particularly when $Q\geq 8$ at high SNR. Similar to the findings in Fig. \ref{NMSE_SNR}(a), TS-PAD-OMP exhibits poor performance with low pilot overhead at relatively high SNR levels, such as $5$ and $15$ dB. However, it demonstrates improved performance when SNR is low, such as $-5$ dB, even with low pilot overhead. 
This can be attributed to its first stage, which facilitates joint angle estimation across all subarrays. It is important to note that since the channel error is determined by the estimation of angle, distance, and path gain parameters, TS-PAD-OMP's sequential estimation approach with error accumulation may lead to performance degradation in various scenarios due to uncertainties.

\begin{figure}
	\centering
	\subfloat[SNR $=-5$ dB]{
		\includegraphics[width=2.09in]{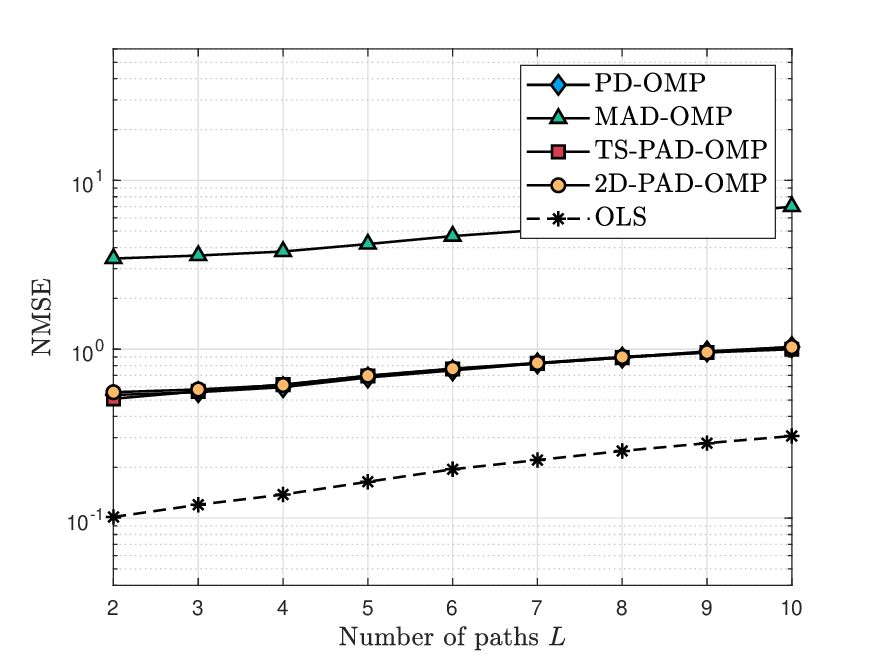}
	}
	\subfloat[SNR $=5$ dB]{
	\includegraphics[width=2.09in]{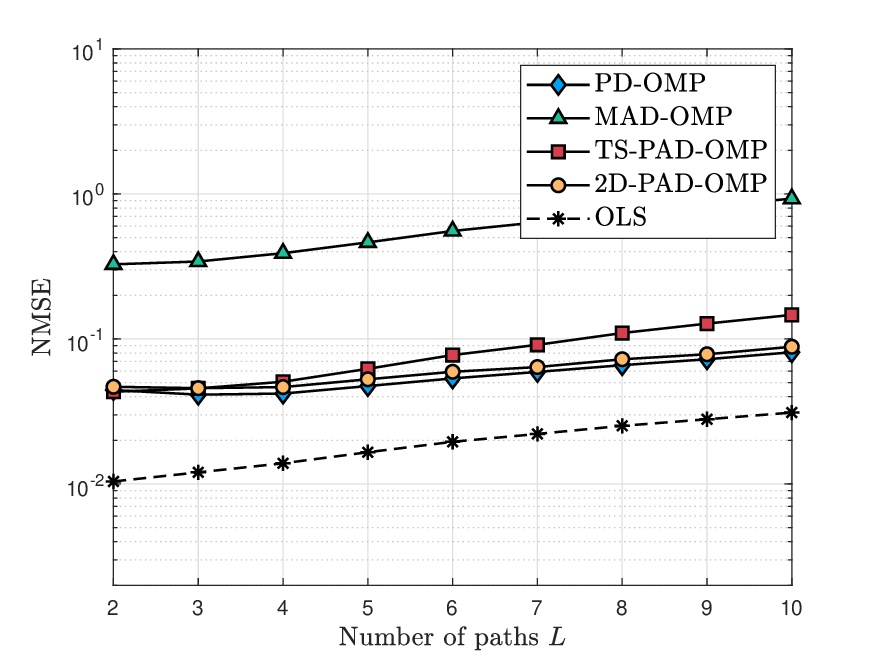}
}   
	\subfloat[SNR $=15$ dB]{
		\includegraphics[width=2.09in]{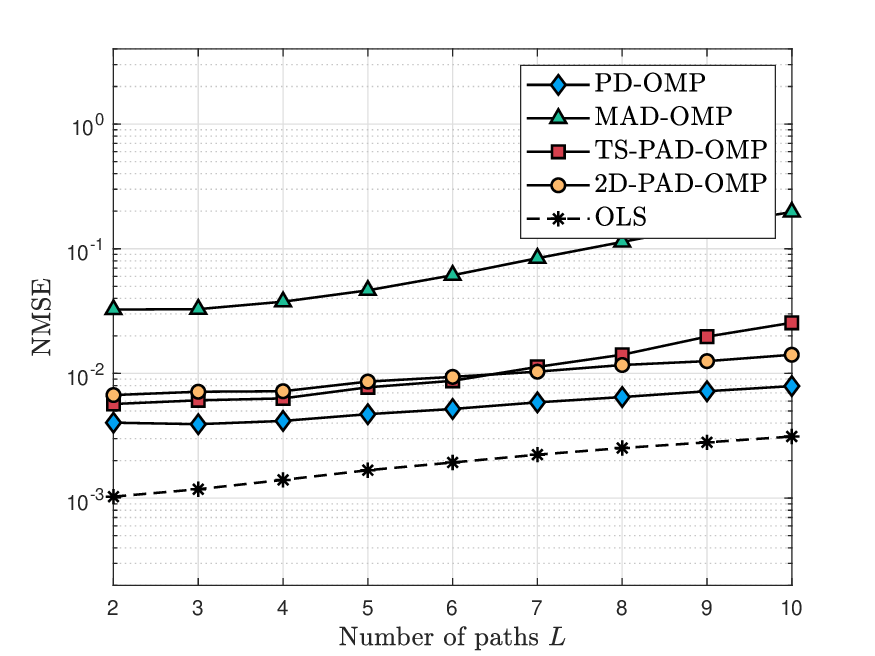}
	}
	\caption{The NMSE performance of different methods varies with $L$ when the SNR is $0$ and $15$ dB.}
	\label{NMSE_L}
\end{figure}

Subsequently, we assess the impact of the number of channel paths $L$ on NMSE performance. In Fig. \ref{NMSE_L}, the NMSE of various methods is plotted for $L$ ranging from $2$ to $10$, with SNR values of $-5$, $5$, and $15$ dB, and $Q=16$. It is observed that the NMSE for all methods increases as the number of paths $L$ rises. This trend is attributed to the growing complexity in the recovery problem, where increased sparsity results in a higher number of parameters needing estimation, thereby diminishing performance. Importantly, as shown in Fig. \ref{NMSE_L}(a), the three methods, PD-OMP, TS-PAD-OMP, and PD-OMP, show comparable performance as $L$ increases, which corresponds to the finding in Fig. \ref{NMSE_SNR}(c). Moreover, as SNR increases, PD-OMP demonstrates its superiority, approaching the lower bound. As a low-complexity implementation, 2D-PAD-OMP achieves comparable performance to PD-OMP in most cases and is therefore suitable for practical usage.

\begin{figure}
	\centering
	\includegraphics[width = 0.49\textwidth]{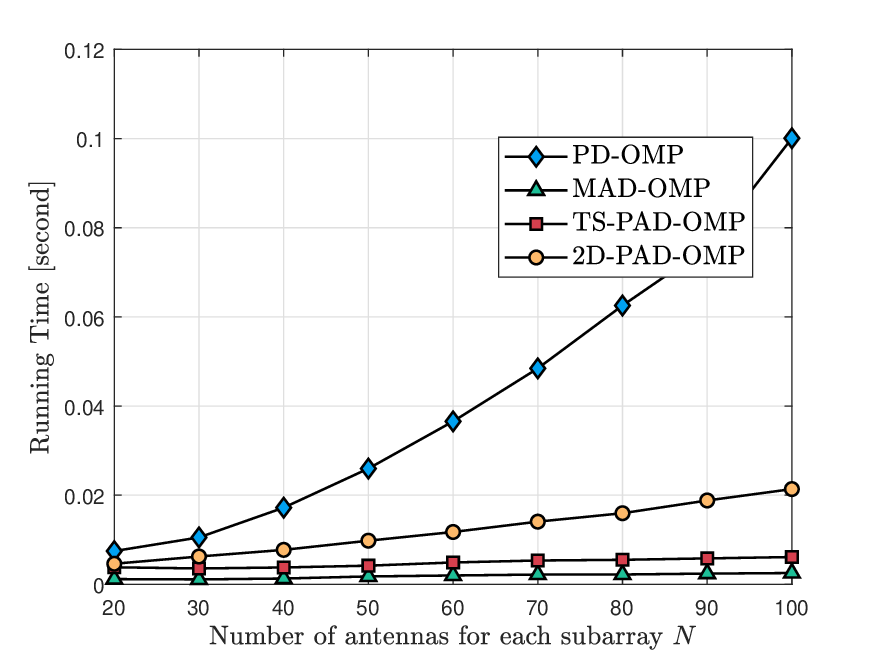}
	\caption{Running time versus the number of antennas for each subarray $N$ for different methods.}
	\label{time}
\end{figure}

\textcolor{red}{In Fig. \ref{time}, we further exhibit the running time on CPU Intel(R) Core(TM) i7-10750H CPU @ 2.60GHz  2.59 GHz for these four methods against the number of antennas for each subarray $N$. Here, the number of subarrays is set to $M=8$, and the number of measurements is set to $Q=20$. It can be observed that PD-OMP exhibits high sensitivity to $N$, making it impractical for high-speed processing requirements in real-world scenarios, especially with larger systems. Conversely, the other three methods demonstrate faster processing times compared to PD-OMP, which aligns with the complexity analysis in Section \ref{CCA}.}
 
\section{Conclusions}\label{Con} 
Towards the conclusion of the paper, we present a concise summary of our research efforts and share some potential directions.
\subsection{Paper Summary}
This paper demonstrates the efficacy of using varied approaches, such as PD-R, MAD-R, and 2D-PAD-R, to model near-field channels with suitable dictionaries. These methods enable a range of channel estimation frameworks, each calibrated to meet specific NMSE or complexity criteria.
Our analysis and simulations assess the complexity of these frameworks, particularly those using OMP. The PD-E framework stands out for its superior NMSE performance, though it comes with the highest complexity. Conversely, TS-PAD-E manages to reduce complexity but at the expense of performance. Notably, 2D-PAD-E achieves performance on par with PD-E but with reduced complexity. Meanwhile, despite its low complexity, MAD-E lags behind in performance.
In summary, 2D-PAD-E, with its 2D dictionary approach, shows considerable promise for THz WSMS systems, balancing complexity and performance effectively. The diversity in near-field channel representation strategies also allows for tailored solutions, whether aiming for low complexity or high performance\textcolor{red}{.}

\subsection{Potential Future Work} 
As this paper explores a relatively simple and idealized scenario for proposing the UCE framework, it presents numerous opportunities for enhancement and further research. Here are some potential avenues for extending our results.
\subsubsection{UPA-MIMO}
The results presented in this paper can be readily extended to UPA systems by estimating both elevation and azimuth angles. Near-field MIMO channel estimation poses certain challenges, as the near-field MIMO line-of-sight component exhibits a much more complex expression compared to MIMO non-line-of-sight. Fortunately, these challenges can be effectively addressed through the cross-field assumption, coupled with the application of the proposed TS-PAD-E and 2D-PAD-E frameworks.

\subsubsection{WSMS-RIS} 
In \cite{SE2}, the study focused on the THz WSMS structure for RISs, where the RIS is divided into multiple widely-spaced sub-RISs. In such scenarios, the pursuit of more efficient channel estimation methods that align with the spatial characteristics of WSMS-RIS holds significant value.

\subsubsection{Off-Grid Recovery}
This paper relies on the on-grid parameter assumption, whereas practical parameters often deviate from the grid. Therefore, the need for off-grid or gridless recovery methods, such as \cite{off1,off2}, becomes apparent for better estimation.
 \subsubsection{Wideband Spatial Effects}
\textcolor{red}{In THz wideband communication systems, wideband spatial effects such as the beam squint effect play a vital role \cite{nf-loc,E2}.} While this aspect is not addressed in this paper, investigating WSMSs with consideration for the near-field beam squint effect is a valuable direction for further research. 
\subsubsection{Distributed XL-Arrays}
Due to potential hardware and complexity challenges with centralized XL-systems, distributed signal processing and cell-free communications have gained significance \cite{Distri1,Distri2}. Based on this, multiple phase-coherent distributed small arrays could be harnessed to create a virtual XL-array. As discussed in Section \ref{DR}, our proposed techniques are well-suited for near-field channel estimation in distributed XL-arrays, with potential for further enhancement.

\bibliographystyle{IEEEtran}
\bibliography{reference.bib}

\vspace{12pt}

\end{document}


\ArticleType{Supplementary File}

\title{Title}{Title for citation}

\author[1]{Aaa AUTHOR}{}
\author[1,2]{Bbb AUTHOR}{{bauthor@xxx.com}}
\author[2]{Ccc AUTHOR}{}
\author[3]{Ddd AUTHOR}{}

\AuthorMark{Author A}

\AuthorCitation{Author A, Author B, Author C, et al}


\address[1]{Affiliation, City {\rm 000000}, Country}
\address[2]{Affiliation, City {\rm 000000}, Country}
\address[3]{Affiliation, City {\rm 000000}, Country}

\maketitle


\begin{appendix}

\section{Importance}
Please use this sample as a guide for preparing your letter. Please read all of the following manuscript preparation instructions carefully and in their entirety. The manuscript must be in good scientific American English; this is the author's responsibility. All files will be submitted through our online electronic submission system at \href{https://mc03.manuscriptcentral.com/scis}{HERE}.

\section{More information}
The examples at the bottom of the .tex file can help you when preparing your manuscript. We are appreciate your effort to follow our style~\cite{1,2}.

\end{appendix}
